\documentclass[%
 aip,numerical,
 jmp,%
 amsmath,amssymb,
 reprint,%
]{revtex4-1}

\usepackage[dvipdfmx]{graphicx}
\usepackage[dvipdfmx]{color}
\usepackage{dcolumn}
\usepackage{bm}
\usepackage{natbib}

\newcommand{\ltsim}{\protect\raisebox{-0.5ex}{$\:\stackrel{\textstyle <}{\sim}\:$}}

\begin{document}

\preprint{AIP/123-QED}

\title[Electrostatic interactions between dust grains]{First-principles simulations of electrostatic interactions between dust grains}

\author{H. Itou}
 \email{h-itou@eps.s.u-tokyo.ac.jp}
\author{T. Amano}
\author{M. Hoshino}%
\affiliation{Department of Earth and Planetary Science, The University of Tokyo}

\date{\today}
\begin{abstract}
We investigated the electrostatic interaction between two identical dust grains of an infinite mass immersed in homogeneous 
plasma by employing first-principles N-body simulations combined with the Ewald method. We specifically tested the possibility of an 
attractive force due to overlapping Debye spheres (ODSs), as was suggested by Resendes et al. (1998). Our simulation results demonstrate 
that the electrostatic interaction is repulsive and even stronger than the standard Yukawa potential. We showed that the measured 
electric field acting on the grain is highly consistent with a model electrostatic potential around a single isolated grain that takes 
into account a correction due to the orbital motion limited theory. Our result is qualitatively consistent with the counterargument 
suggested by Markes and Williams (2000), indicating the absence of the ODS attractive force.
\end{abstract}
\maketitle


\section{Introduction}
Dust grains are quite common in astrophysical environments. They are thought to exist in, for example, interstellar molecular clouds, 
protoplanetary disks, planetary rings, the Earth's magnetosphere, and tails of comets. In addition, in laboratories, the lattice 
formation of dust grains, known as Coulomb crystallization, is a well-known phenomenon that has fascinated many researchers. Dust grains 
immersed in plasmas usually acquire a large amount of charge through several charging processes, such as collisions with plasma particles 
and photoemission. Such charged grains and the ambient plasma are electromagnetically coupled with each other, forming so-called dusty 
plasmas or complex plasmas. Dusty plasma has been studied for both industrial and astrophysical applications, largely motivated by the 
in-situ detection of dust grains in the Solar System and Ikezi's prediction, and subsequent experimental verification of Coulomb 
crystallization.\cite{Goertz89,Angelis92,Shukla09,Shukla01,text,Ikezi86}

When collisions between dust grains and plasma particles are dominant among the charging processes, dust grains become negatively charged 
because the thermal velocity of electrons is generally higher than that of ions, resulting in a larger electron current. Therefore, one 
would expect a repulsive shielded electrostatic Coulomb potential (or Yukawa potential) to exist. In reality, 
however, forces acting on dust grains may be much more complex because the interaction forces between charged dust grains are mediated by 
the ambient plasma in a complicated manner. There has been much discussion on forces acting between dust grains, including attractive 
forces for which the ambient plasma response plays the essential role.\cite{Shukla09,Lampe00} It is necessary to understand the nature of 
such attractive interactions among dust grains because they may play a role in the aggregation or crystallization of dust grains observed 
in laboratories, as well as the formation of stars and planets in the dense cores of interstellar molecular clouds.

One such attractive force acting between two grains, and on which we focus in the present study, is the force due to overlapping Debye 
spheres (ODSs).\cite{Resendes98} According to Resendes et al. (1998), when two charged dust grains (each having charge $q$) exist in a 
plasma, their interaction potential, including the electrostatic energy of ambient plasma particles, may be modified from the simple 
Yukawa potential. The potential in this case may be written as
\begin{equation}
	\label{l-j}
	q{\phi_{{\rm ODS}}}\left(d\right)=\frac{q^{2}}{\lambda_{{\rm D}}}\left(\frac{\lambda_{{\rm D}}}{d}-\frac{1}{2} \right)\exp\left(-\frac{d}{\lambda_{{\rm D}}} \right)+{\rm constant},
\end{equation}
where $\lambda_{{\rm D}}$ is the Debye length and $d$ is the intergrain distance. This is similar to the Lennard-Jones potential, which 
is repulsive at short distances and weakly attractive at longer distances. It is clear that a Lennard-Jones-like potential can assist the 
processes of aggregation and crystallization, and in fact, it has been shown that the attractive force due to ODSs has a drastic effect on 
aggregation and crystallization in dusty plasmas if indeed effective.\cite{Hou09} It has also been suggested that the ODS attractive 
force may enhance gravitational instability and assist the formation of stars and planets in astrophysical environments.\cite{Shukla06} 
On the other hand, the derivation of this attractive potential is based on several nontrivial assumptions that need to be verified. For 
instance, one must assume that the electrostatic potential around a dust grain is given by the Yukawa potential:
\begin{equation}
	\label{yukawa}
	q{\phi}\left(r\right)=\frac{q^{2}}{r}\exp\left(-\frac{r}{\lambda_{{\rm D}}}\right).
\end{equation}
In addition, linear superposition of the potential around two dust grains (with ODSs) should be valid in order for such an attractive 
force to exist. Since the concept of Debye shielding is the key to understanding the attractive force, one must be careful in adopting 
these assumptions. Furthermore, the derivation of the ODS attractive ${\it force}$ from Eq. (\ref{l-j}) assumes that the force 
operating between the grains is given by the derivative of Eq. (\ref{l-j}) with respect to the intergrain distance $d$. We note that 
Markes and Williams (2000) pointed out that this assumption is incorrect in that it does not take into account energy exchange with the 
ambient plasma.\cite{Markes00} Lampe et al. (2000) also suggested 
that, on the basis of orbital motion limited (OML) theory, such an attractive force would not exist.\cite{Lampe00} Nevertheless, those 
counterarguments are also based on some non-trivial assumptions. Consequently, the 
existence or nonexistence of the ODS attractive force has yet remained a controversial issue. 

The purpose of our study is thus to investigate the validity of the theory of the ODS attractive force from first principles. We employ 
the direct N-body simulation method in which all particle-particle interactions acting through the electrostatic 
Coulomb force are calculated. This first-principles approach allows us to investigate the electrostatic potential structure of 
sub-Debye scales without making any assumptions, and thus provides a direct answer to the problem. 

It is demonstrated herein that the electric field acting on a charged grain actually deviates from the standard 
Yukawa-type field in general. We find that the electrostatic force acting between two dust grains is repulsive rather than attractive, 
which may be well explained by OML theory for an isolated test charge. There is no noticeable signature of the net attractive force due 
to the 
effect of ODSs around dust grains. Our result is qualitatively consistent with the analysis given by Markes and Williams 
(2000). Although the simulations were performed within a limited range of plasma parameters, this strongly indicates the ODS attractive 
force is absent in reality.

\section{Simulation method}
Our N-body simulations are performed in a periodic system (surrounded by a virtual perfectly conducting medium at the infinite distance). 
The system consists of the 
simulation box and its replicas, and the box contains many plasma particles (ions and electrons) and two charged dust grains. For the 
time integration, the Coulomb force acting on each particle must be evaluated by taking the summation over all particles. Since the 
Coulomb interaction is a long-range interaction, convergence of the summation is very slow and the calculation of contributions from many 
particles at long distances significantly increases the number of operations required.

We thus adopt the Ewald method, which allows us to accelerate the summation by dividing it into two parts: one in real space and the 
other in wavenumber space. For instance, the electrostatic potential may be calculated as follows:
\begin{equation}
	\label{ewald}
	U=U_{{\rm real}}+U_{{\rm wave}}-U_{{\rm self}},
\end{equation}
\begin{equation}
	\label{real}
	U_{{\rm real}}=\frac{1}{2}\sum_{i,j}\sum_{n}{\frac{q_{i}q_{j}}{r_{ijn}} {\rm erfc}\left(\frac{r_{ijn}}{\sigma} \right)},
\end{equation}
\begin{equation}
	\label{wave}
	U_{{\rm wave}}=\frac{1}{2}\sum_{i,j}\sum_{{\bm k}\neq 0}{q_{i}q_{j}\frac{\exp\left[{-\pi^{2}\sigma^{2}k^{2}+2\pi i{\bm k}\cdot\left({\bm r}_{i}-{\bm r}_{j}\right)}\right]}{\pi V k^{2}}},
\end{equation}
\begin{equation}
	\label{self}
	U_{{\rm self}}=\frac{1}{\sqrt{\pi}\sigma}\sum_{i}{q_{i}^{2}}.
\end{equation}
Here, $n$ represents the labels of boxes, $r_{ijn}$ is the distance between particles $i$ and $j$ in box $n$, $q_{i}$ is the charge of 
particle $i$, ${\bm k}$ is the wavenumber vector, and $V$ is the volume of the box. The parameter $\sigma$ gives a cut-off radius beyond 
which the direct summation in real space, Eq. (\ref{real}), is replaced by that in wavenumber space, Eq. (\ref{wave}). Note that in Eq. 
(\ref{real}), the term $n=0$ has to be excluded for $i=j$. This method approximates long-wavelength modes associated with the long-range 
nature of the Coulomb interaction in wavenumber space with the aid of the Fourier transform, whereas short-wavelength components arising 
from close encounters between particles are accurately calculated. The electric field is given by the spatial derivatives of Eqs. 
(\ref{real}) and (\ref{wave}) and is calculated in the same way.\cite{Deserno98-1,Pollock96}

In calculating Eq. (\ref{real}), we introduce a small softening parameter $\epsilon$ and rewrite Eq. (\ref{real}) as
\begin{equation}
	\label{ereal}
	U_{{\rm real}}=\frac{1}{2}\sum_{i,j}\sum_{n}{\frac{q_{i}q_{j}}{\sqrt{r_{ijn}^{2}+\epsilon^{2}}} {\rm erfc}\left(\frac{\sqrt{r_{ijn}^{2}+\epsilon^{2}}}{\sigma} \right)}.
\end{equation}
With the softening technique, we ignore large-angle scatterings between particles at distances $\ltsim \epsilon$ because resolving such 
scatterings would require very small time steps. Since we are interested in weakly coupled space and astrophysical plasmas that are 
defined by a large plasma parameter $\Lambda$ (where small-angle scatterings play the dominant role),  we think this technique is 
reasonable for our purpose.

Having calculated the electric fields acting on particles, we can solve the equations of motion for each particle:
\begin{equation}
	\label{newton1}
	m_{i}\frac{d}{dt}{\bm v}_{i}=q_{i}{\bm E},
\end{equation}
\begin{equation}
	\label{newton2}
        \frac{d}{dt}{\bm r}_{i}={\bm v}_{i},
\end{equation}
where $m_{i}$, ${\bm v}_{i}$, and ${\bm r}_{i}$ are the mass, velocity, and position of particle $i$, respectively, and ${\bm E}$ is the 
electric field at each particle position ${\bm r}_{i}$. In Eq. (\ref{newton1}), assuming nonrelativistic plasma temperatures, 
$v_{i}/c\ll1$, we ignore the effect of magnetic fields.

Throughout the present paper, the masses of ions and electrons are assumed to be equal to allow the system to relax quickly to an 
equilibrium state. This assumption may be justified because the mass ratio affects only the time scale, and structures of the equilibrium 
state can be assumed to be independent of the mass ratio. Therefore, we only discuss the properties of equilibrium states. Note that 
because of the symmetry of ion and electron masses, the sign of the grain charge is irrelevant. Simulations are performed with two 
identical dust grains of infinite mass in the box. That is to say, the grain mass is so large that the change in positions can be ignored 
on the simulation time scale, which is typically limited to a few plasma oscillation periods. The effect of finite grain size is also 
ignored. These assumptions are made to simplify the problem as 
much as possible for our purpose of investigating the electrostatic interactions between plasma particles and dust grains.

\section{simulation result}
Simulations were initialized with plasma particles distributed randomly in space, and two dust grains placed at fixed distances in the 
box. The velocity distribution was initialized to a Maxwellian distribution for a given temperature. Time integration was carried out until 
the system reached an equilibrium state, at which point we measured the properties of the system. The simulation box was a cuboid whose 
dimensions were $2L$ in the $x$ direction and $L$ in the $y$ and $z$ directions. Throughout this paper, we use a softening parameter of 
$\epsilon=0.03L$ in simulations. Each grain was located at $\left(y,z\right)=\left(L/2,L/2\right)$, and the intergrain distance along the 
$x$ axis was varied in each simulation run. By comparing the equilibrium states of different runs, we measured the dependence on the 
intergrain distance.
\begin{center}
	\begin{figure}[t]
		\includegraphics[width=90mm]{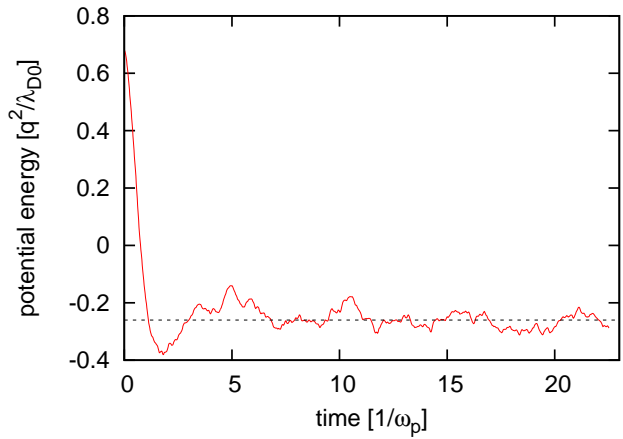}
		\caption{Temporal evolution of total electrostatic potential energy for a run with $d=0.1L$, $q=1000e$, 
                $2L^{3}n_{{\rm e}}=10000$, $2L^{3}n_{{\rm p}}=8000$, $\lambda_{{\rm D}0}\simeq0.11L$, $\lambda_{{\rm D}}\simeq0.12L$, and 
                $\Lambda\simeq16$. The dotted line indicates the equilibrium value.}
		\label{ene}
	\end{figure}
\end{center}

The system is characterized by the dust charge $q$ and the intergrain distance $d$. The number densities of electrons and ions are 
denoted $n_{{\rm e}}$ and $n_{{\rm p}}$, which are chosen so that charge neutrality (including dust charges) is satisfied. In the 
following, time and space are respectively normalized by the inverse plasma frequency $1/\omega_{p}$, where 
$\omega_{p}=\left(4\pi n_{{\rm e}} e^{2}/m_{{\rm e}}+4\pi n_{{\rm p}} e^{2}/m_{{\rm p}}\right)^{1/2}$, and the Debye length 
$\lambda_{{\rm D}}$. Note that the Debye length is defined as 
$\lambda_{{\rm D}}=\left(4\pi n_{{\rm e}} e^{2}/kT_{{\rm e}}+4\pi n_{{\rm p}} e^{2}/kT_{{\rm p}}\right)^{-1/2}$, including both ion and 
electron contributions, and the temperatures of the resultant equilibrium states are used. Here, $e$ is the elementary charge, and 
$m_{{\rm e}}$, $m_{{\rm p}}$, $T_{{\rm e}}$, and $T_{{\rm p}}$ are the electron mass, proton mass, electron temperature, and proton 
temperature, respectively. Note that we always assumed that the initial electron and proton temperatures were the same for simplicity. 

In Fig.\ref{ene}, the time variation of the potential energy integrated over the simulation box is shown for the example of a run with an 
intergrain distance of $d=0.1L$. The energy is normalized by $q^2/\lambda_{{\rm D}0}$, where $\lambda_{{\rm D}0}$ is the Debye length 
defined by the initial temperature. In this run, $q=1000e$, $2L^{3}n_{{\rm e}}=10000$, $2L^{3}n_{{\rm p}}=8000$, 
$\lambda_{{\rm D}0}\simeq0.11L$, $\lambda_{{\rm D}}\simeq0.12L$, and 
$\Lambda\equiv\left(n_{{\rm e}}+n_{{\rm p}}\right)\lambda_{{\rm D}}^{3}\simeq16$. Generally speaking, the Debye length in the final 
equilibrium state, denoted $\lambda_{{\rm D}}$, actually differs from $\lambda_{{\rm D}0}$, as explained below. We see from Fig.\ref{ene} 
that the potential energy decreases during the first $\sim1/\omega_{p}$, and then fluctuates around the equilibrium value. This initial 
decrease in the potential may be explained by the redistribution of plasma particles due to Debye shielding. This decrease in the 
potential energy is compensated by an increase in the plasma temperature, changing the Debye length from the initial value accordingly. 
All runs discussed in this paper showed essentially the same trend. We thus assume that the equilibrium was achieved by the time 
$\omega_{{\rm p}}t\sim8$, and physical quantities averaged after this time were regarded as equilibrium values.
\begin{center}
	\begin{figure}[t]
		\includegraphics[width=90mm]{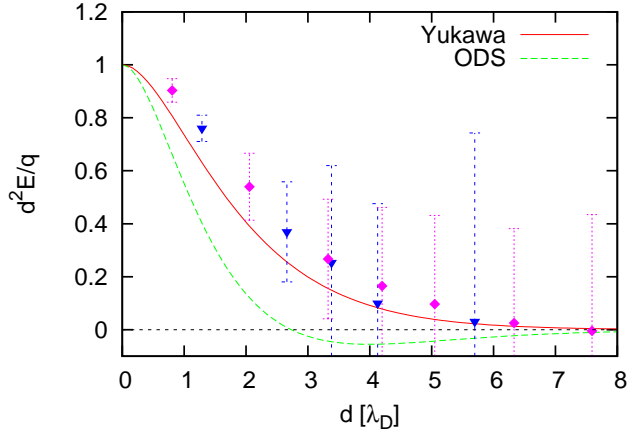}
		\caption{Summary of simulation results. The normalized electric field acting on the grain multiplied by $d^{2}$ is shown 
                as a function of the intergrain distance $d$. Note that the distance is normalized by the Debye length defined with the 
                kinetic energy measured at the equilibrium states rather than the initial temperature. 
                The red and green lines are the theoretical curves expected from the 
                standard Yukawa potential and the ODS attractive potential, respectively. Blue triangles and magenta diamonds show the 
                results of our simulations with $\Lambda\simeq13$ and $\Lambda\simeq16$, respectively.}
		\label{res_ele}
	\end{figure}
\end{center}

Figure \ref{res_ele} summarizes the results of our simulations. Blue triangles show the results for $2L^{3}n_{{\rm e}}=5000$, 
$2L^{3}n_{{\rm p}}=3000$, $\lambda_{{\rm D}}/L\simeq0.15L$, and $\Lambda\simeq13$. Individual triangles represent intergrain distances of 
$d=0.2L, 0.4L, 0.5L, 0.6L, 0.8L$. Simulations with a different set of parameters ($2L^{3}n_{{\rm e}}=10000$, 
$2L^{3}n_{{\rm p}}=8000$, $\lambda_{{\rm D}}/L\simeq0.12L$, and $\Lambda\simeq16$) were also run, and the results are shown by magenta 
diamonds; in this case, the intergrain distances were $d=0.1L, 0.25L, 0.4L, 0.5L, 0.6L, 0.75L, 0.9L$. In all runs, $q=1000e$ and 
$\lambda_{{\rm D}0}\simeq0.11L$. Note that $\lambda_{{\rm D}}$, which normalizes the intergrain distances in Fig.\ref{res_ele}, was 
defined at the equilibrium states, and thus not 
necessarily the same in each run because the self-consistent increase in temperature depends on plasma densities, plasma parameters and intergrain 
distances $d$. The red and green lines in Fig.\ref{res_ele} show the theoretical curves expected from the standard Yukawa potential and 
the ODS attractive potential of Resendes et al. (1998), respectively, which are written as
\begin{equation}
	\label{yukawa-e}
        qE_{{\rm Yukawa}}\left(d\right)=\frac{q^{2}}{\lambda_{{\rm D}}^{2}}\left[\left(\frac{\lambda_{{\rm D}}}{d}\right)^{2}+\frac{\lambda_{{\rm D}}}{d}\right]\exp\left(-\frac{d}{\lambda_{{\rm D}}}\right)
\end{equation}
and
\begin{equation}
	\label{l-j-e}
        qE_{{\rm ODS}}\left(d\right)=\frac{q^{2}}{\lambda_{{\rm D}}^{2}}\left[\left(\frac{\lambda_{{\rm D}}}{d}\right)^{2}+\frac{\lambda_{{\rm D}}}{d}-\frac{1}{2}\right]\exp\left(-\frac{d}{\lambda_{{\rm D}}}\right).
\end{equation}
Eq. (\ref{l-j-e}) assumes that the force on the grain is given by the derivative of Eq. (\ref{l-j}) with respect to $d$. The error bars 
represent the standard deviation ($1\sigma$) of temporal fluctuations after the system has reached an equilibrium state. Note that when calculating 
the electric field acting on the grain, we used a softening parameter of $\epsilon=d/12$, which is different from that used in the 
simulation to reduce the variance of the measured electric fields. That is to say, the softening parameter $\epsilon$ is chosen to be 
proportional to the intergrain distance, whereas it is constant in all simulations. This choice is mainly motivated by the conjecture 
that the equilibrium electrostatic structure will not strongly depend on the softening parameter. However, some remarks must be made 
before discussing the results.

First, the effect of softening is not seen even at $d \lesssim 0.03 L\left(\simeq0.2-0.25\lambda_{{\rm D}}\right)$ because the softening 
parameters used in the calculations are smaller than the simulation value at $d < 0.36L\simeq2.5-3\lambda_{{\rm D}}$. In the region where 
the softening effect is significant, it is obvious that the potential approaches the Coulomb potential because the softening parameter in the 
simulations is chosen to be smaller than the mean particle distance. Therefore, this will not change our conclusions.

Second, the error bars may be underestimated at $d > 0.36L$ because the softening parameter used in the calculation becomes larger than 
that in the simulations. (Note that large error bars are caused by close encounters with plasma particles.) In any case, the error bars 
are so large that it is difficult to extract a physically meaningful argument in this regime.

Third, we have confirmed that calculation with a constant softening parameter of $\epsilon=0.03L$ (i.e., consistent with the simulations) 
does not change the result substantially. Although the error bars in the far regions, i.e., $d > 0.36L$, tend to increase, the average 
electric fields stay within the error bars shown in Fig.\ref{res_ele}.

Based on these discussions, we believe that the simulation results are reliable at least in the intermediate regime, i.e., 
$1\lambda_{{\rm D}}\lesssim d\lesssim2.5\lambda_{{\rm D}}$. In this region, it is evident from Fig.\ref{res_ele} that the simulation 
results deviate from the theoretical prediction of the ODS attractive potential beyond $2\sigma$. The result also suggests that 
the electric fields acting on 
the grain are even larger than the standard Yukawa potential prediction. Although the large error bars make it difficult to draw 
conclusions from this result alone, the systematic deviation from the theoretical predictions suggests that the underlying assumptions 
made in the derivation of (\ref{yukawa-e}) and (\ref{l-j-e}) may be violated. In the next section, we discuss possible reasons for this 
discrepancy between the theory and simulations.

\section{Discussion}
Our simulation results show that the force between two dust grains is repulsive and stronger than that predicted by the standard Yukawa 
potential Eq. (\ref{yukawa}). At first, we discuss the validity of Eq. (\ref{yukawa}). When the grain radius is negligible, the 
functional form of the Yukawa potential itself must be correct at large distances, where the shielding is nearly complete and the 
first-order expansion of the Boltzmann-type density distribution is appropriate. In fact, Poisson's equation and the linearized Boltzmann 
distributions give 
\begin{equation}
	\label{long}
        q\phi\left(r\right)=\alpha\frac{q^{2}}{r}\exp{\left(-\frac{r}{\lambda_{{\rm D}}}\right)}.
\end{equation}
However, the coefficient $\alpha$ (integration constant) in Eq. (\ref{long}) is unknown and must be determined by the inner boundary 
condition. In standard textbooks, it is determined by assuming that the outer solution smoothly connects to the bare Coulomb potential at 
$r \rightarrow 0$, which gives $\alpha=1$.

On the other hand, according to OML theory, $\alpha\neq1$ in general. In OML theory, when particle absorption by dust grains is ignored, 
the density distribution of ions around a negatively charged dust grain may be written as \cite{Lampe00} 
\begin{equation}
	\label{oml}
	n_{{\rm p}}=n_{0}\left[\exp\left(-\frac{e\phi}{kT_{{\rm p}}}\right){\rm erfc}\left(\sqrt{-\frac{e\phi}{kT_{{\rm p}}}} \right)+\frac{2}{\sqrt{\pi}}\sqrt{-\frac{e\phi}{kT_{{\rm p}}}} \right]
\end{equation}
instead of the Boltzmann distribution
\begin{equation}
	\label{boltz-p}
	n_{{\rm p}}=n_{0}\exp\left(-\frac{e\phi}{kT_{{\rm p}}}\right),
\end{equation}
whereas the electron density distribution is written as
\begin{equation}
	\label{boltz-e}
	n_{{\rm e}}=n_{0}\exp\left(\frac{e\phi}{kT_{{\rm e}}}\right)
\end{equation}
in both cases. It is easy to show that Eqs. (\ref{oml}) and (\ref{boltz-p}) give the same dependence on $e\phi/kT$ when expanded to first 
order in $e\phi/kT\ll1$, meaning that the functional form is the same far from the grain.

Since the OML correction given by Eq. (\ref{oml}) gives an ion density much lower than that suggested by the Boltzmann distribution given 
by Eq. (\ref{boltz-p}) close to the grain, the shielding of the potential becomes weaker. We may thus expect $\alpha\geq1$ in general if 
the OML correction is taken into account.\cite{Lampe00} The parameter $\alpha$ may be determined by the solution in the inner region, 
where the OML correction may become important. On the other hand, the OML solution must also be connected to the bare Coulomb potential 
\begin{equation}
	\label{short}
        q\phi\sim \frac{q^{2}}{r},
\end{equation}
at distances on the order of the mean interparticle distance $a$, which is defined as 
\begin{equation}
	\label{wig2}
        \frac{a}{\lambda_{{\rm D}}}\equiv \sqrt[3]{\frac{3}{4\pi \Lambda}},
\end{equation}
where $\Lambda$ is the plasma parameter. While it is difficult to analyze the potential structure analytically in the inner region with 
the OML correction, we expect $\alpha\left(\Lambda\right)$ to be a decreasing function of $\Lambda$ because a larger $\Lambda$ narrows 
the region in which the OML correction should be taken into account and strengthens the shielding effect.
\begin{center}
	\begin{figure}[t]
		\includegraphics[width=90mm]{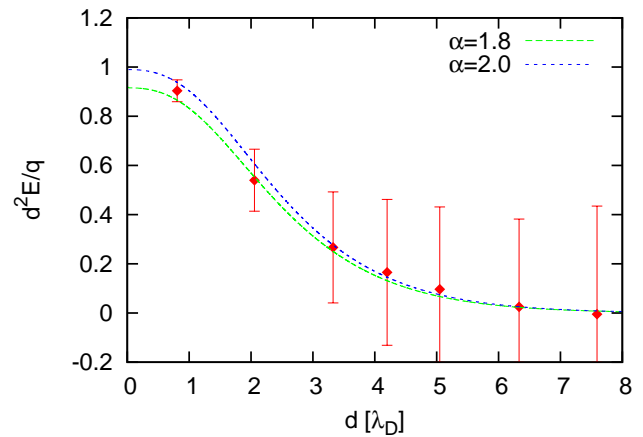}
		\caption{Comparison between simulation results and theoretical models including the OML correction for the electric field 
                acting on the grain. Only results with $\Lambda\simeq16$ are shown.}
		\label{com}
	\end{figure}
\end{center}

To determine the value of $\alpha$, Figure \ref{com} compares the simulation results and a 
theoretical electric field around ${\it a}$ ${\it single}$ ${\it isolated}$ ${\it grain}$ including the OML correction. That is to say, 
the potential $\phi$ was determined by solving Poisson's equation, 
\begin{equation}
	\label{poi}
        \nabla^{2}\phi=-4\pi e\left(n_{{\rm p}}-n_{{\rm e}}\right),
\end{equation}
with the ion and electron densities given by Eqs. (\ref{oml}) and (\ref{boltz-e}), respectively. $n_{0}$ in Eqs. (\ref{oml}) and 
(\ref{boltz-e}) was approximated as $n_{0}=(n_{{\rm e}0}+n_{{\rm p}0})/2$ for simplicity. The plasma parameter was $\Lambda\simeq16$, 
which is almost the same as that in our simulations. The electric field $E$ was calculated by taking spatial derivatives of $\phi$. As we 
have already mentioned, the functional form of Eq. (\ref{long}) should be valid far from the grain even if the OML correction is 
included. Therefore, Poisson's equation was integrated from a large radial distance toward the inner region by taking $\alpha$ as a free 
parameter. We then tried to find the values of $\alpha$ for which this theoretical solution reasonably matched the simulation results. It 
is readily seen from Fig.\ref{com} that the simulation results are well explained by this model with $\alpha\simeq 1.8-2.0$. Note again 
that the theoretical curve is for an isolated grain, whereas the simulation results are obtained with two dust grains. This means that 
the effect of ODSs is not observed, at least to a detectable level beyond the error bars of our simulations. This result is qualitatively 
consistent with the suggestion by Markes and Williams (2000). They have shown explicitly that the electrostatic force acting between two 
grains surrounded by a plasma is repulsive by solving Poisson's equation. The critical assumption in their model is that the ion and 
electron densities can be written as a function of the local electrostatic potential alone. Although this assumption sounds reasonable 
for instance in the collisionless limit where OML theory should apply, its validity must be tested carefully. On the other hand, our first 
principles approach free from such an assumption also demonstrates a repulsive nature for the electrostatic interaction. Furthermore, the 
fact that the electric field around the grain is consistent with the OML theory indicates the assumption made by Markes and Williams 
(2000) is indeed reasonable.

\begin{center}
	\begin{figure}[t]
		\includegraphics[width=90mm]{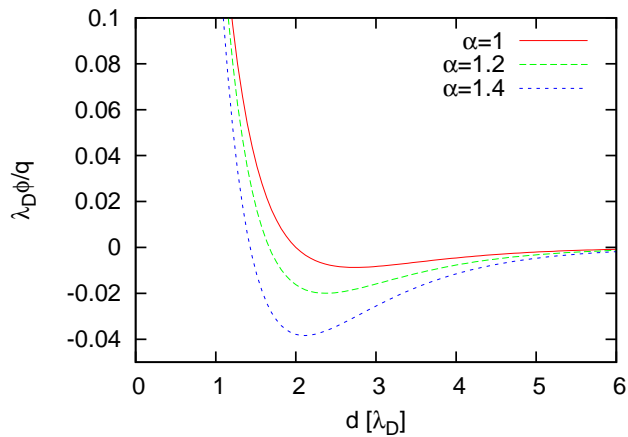}
		\caption{Modified ODS attractive potential given by Eq. (\ref{ml-j}). Red, green, and blue lines represent 
                $\alpha = 1, 1.2, 1.4$, respectively.}
		\label{mod}
	\end{figure}
\end{center}

One might argue that the fact that $\alpha\neq1$ explains the discrepancy between the simulation results and ODS theory, but this is not 
the case. Assuming that linear superposition of the potential is also possible for $\alpha\neq1$, we can easily calculate the ODS 
attractive force for this case as well. The resulting attractive potential force may be written as 
\begin{equation}
	\label{ml-j}
	q{\phi_{{\rm ODS}}}\left(d\right)=\alpha\frac{q^{2}}{\lambda_{{\rm D}}}\left(\frac{\lambda_{{\rm D}}}{d}-\frac{\alpha}{2}\right)\exp\left(-\frac{d}{\lambda_{{\rm D}}} \right),
\end{equation}
which is shown in Fig.\ref{mod} for $\alpha=1, 1.2, 1.4$. It can be easily understood that the potential minimum moves inward and the 
depth increases as $\alpha$ increases. In fact, an easy analytical calculation confirms this tendency. Clearly, $\alpha\neq1$ does not 
help to explain the discrepancy.

Although it is not easy to analytically determine the value of $\alpha$ in general, we can estimate the upper and lower bounds as 
follows. We define $r_{{\rm c}}$ as a solution to the equation 
\begin{equation}
	\label{exp-equ}
        \frac{\alpha}{4\pi\Lambda}\frac{q}{e}\frac{\lambda_{{\rm D}}}{r_{{\rm c}}}\exp{\left(-\frac{r_{{\rm c}}}{\lambda_{{\rm D}}}\right)}=1,
\end{equation}
where the left-hand side is the normalized outer potential. An analytic solution to this equation is given by
\begin{equation}
	\label{sol}
	r_{{\rm c}}=\lambda_{{\rm D}} W\left(\frac{\alpha}{4\pi \Lambda}\frac{q}{e}\right),
\end{equation}
where $W\left(x\right)$ is the inverse function of $x=W\exp\left(W\right)$, which is also known as the Lambert W-function. The potential 
at $r=r_{{\rm c}}$ may be approximated by $\phi\left(r_{{\rm c}}\right)=\alpha q \exp{\left(-r_{{\rm c}}/\lambda\right)}/r_{{\rm c}}$ and 
should be bounded by $q \exp{\left(-a/\lambda\right)}/a$ and the bare Coulomb potential $q/a$, leading to the inequality 
$1\leq\alpha\leq \exp{\left(r_{{\rm c}}/\lambda\right)}$. Using $\Lambda$ and $q/e$, we can rewrite this inequality as
\begin{equation}
	\label{ine1}
        1\leq\alpha\leq \exp{\left(\frac{1}{4\pi \Lambda} \frac{q}{e}\right)}.
\end{equation}
This estimate must be modified when $\Lambda$ is much larger than the critical value $\Lambda_{{\rm c}}$ for which the condition 
$a=r_{{\rm c}}$ is satisfied. When $a\gg r_{{\rm c}}$, $\phi\left(a\right)$ rather than $\phi\left(r_{{\rm c}}\right)$ must be used for a 
similar comparison, yielding
\begin{equation}
	\label{ine2}
        1\leq\alpha\leq \exp{\left(\sqrt[3]{\frac{3}{4\pi\Lambda}}\right)}.
\end{equation}
The condition $r_{{\rm c}}=a$ leads to $\Lambda_{{\rm c}}\sim\left(q/e\right)^{3/2}$, which can also be expressed as 
$kT_{{\rm c}}\sim eq/a$ with a critical temperature $T_{{\rm c}}$. From this, it is clear that when the temperature is above the critical 
value, the plasma is weakly coupled even with dust grains having relatively large charge.  This indicates that the OML correction in this 
regime is only a minor modification, and essentially, the Yukawa-type potential in the far zone directly connects to the bare Coulomb 
potential. 

In our simulations, since we used large dust charges with relatively small numbers of particles, the plasma parameter is smaller than the 
critical value. Note that the plasma parameter of dusty plasmas in space is usually huge, and so is almost always above the critical 
value. Our choice of dust charge was motivated by the fact that the theoretical ODS attractive force is proportional to $q$, and the 
effect is expected to be more pronounced for larger dust charges. As a drawback, we were forced to use sub-critical plasma parameters 
owing to limited computational resources. Because of this, it was not possible to draw a final conclusion. Nevertheless, the qualitative 
consistency between our results and the counterargument against the ODS attractive force strongly indicates that the ODS attractive force 
may not operate in reality. In particular, we believe the assumption that the derivative of the potential energy of the whole system with 
respect to the intergrain distance provides a net force acting on the grain is incorrect as was pointed out by Markes and Williams 
(2000). Equation (\ref{ine2}) shows that, when the plasma parameter is sufficiently large, $\alpha$ becomes almost unity and the 
potential structure around the grain approaches Eq. (\ref{yukawa}), on which the derivation of the ODS attractive potential is based. 
Even in this parameter regime, our results suggest that the electric field acting on the grain is given by the spatial derivative of the 
potential at the grain's position rather than that of the potential of the whole system with respect to the intergrain distance. In this 
case, the electrostatic force acting between two dust grains is always repulsive.

Of course, our results should apply only to the simplest situation where two infinitely small dust grains remain at rest with respect to 
an ambient fully ionized collisionless plasma. There has been a lot of discussion on the force acting on dust grains that may be affected 
by, e.g., finite grain size, relative streaming between the plasma and grains. Comprehensive understanding of the net force due the 
combined effect of those contributions is needed for, e.g., star and planet formation in astrophysical environments.

\section{Conclusion}
We investigated the electrostatic interaction between dust grains surrounded by a plasma by employing first-principles N-body simulations combined with 
the Ewald method. It was shown that the interaction between two 
charged dust grains is repulsive and its magnitude is somewhat larger than that derived from the Yukawa potential. The force acting 
on the dust grains was explained by OML theory for a single isolated grain quite well. The result is consistent with the analysis given 
by Markes and Williams (2000). Consequently, we think that the electrostatic force acting between dust grains are always repulsive. 
Nevertheless, since our simulations have been performed only in a limited parameter range, a final conclusion awaits 
simulations with much higher plasma parameters, which will be made possible by adopting modern numerical schemes such as 
particle-particle particle-mesh and special-purpose GRAPE (GRAvity-piPE) computers for N-body simulations.
\cite{text2,Yamamoto06}

\section{Acknowledgement}
We are grateful to the anonymous referee for his/her critical and constructive comments on the manuscript.

%

\end{document}